\pdfoutput=1
\documentclass{PoS}

\title{Impact of Neutrinoless Double Beta Decay\\ on Models of Baryogenesis}

\ShortTitle{Impact of Neutrinoless Double Beta Decay on Models of Baryogenesis}

\author{\speaker{Julia Harz}\\
        Sorbonne Universit\'es, Institut Lagrange de Paris (ILP), 98 bis Boulevard Arago, 75014 Paris, France\\
Sorbonne Universit\'es, UPMC Univ Paris 06, UMR 7589, LPTHE, F-75005, Paris, France\\
CNRS, UMR 7589, LPTHE, F-75005, Paris, France \\
Department of Physics and Astronomy, University College London, London WC1E 6BT, United Kingdom\\
        E-mail: \email{jharz@lpthe.jussieu.fr}}

\author{Frank F. Deppisch\\
       Department of Physics and Astronomy, University College London, London WC1E 6BT, United Kingdom}
       
\author{Wei-Chih Huang\\
       Fakult\"at f\"ur Physik, Technische Universit\"at Dortmund, 44221 Dortmund, Germany\\
       Department of Physics and Astronomy, University College London, London WC1E 6BT, United Kingdom}

\abstract{Interactions that manifest themselves as lepton number violating processes at low energies in combination with sphaleron transitions typically erase any pre-existing baryon asymmetry of the Universe. We demonstrate in a model independent approach that the observation of neutrinoless double beta decay would impose a stringent constraint on mechanisms of high-scale baryogenesis, including leptogenesis scenarios. Further, we discuss the potential of the LHC to model independently exclude high-scale leptogenesis scenarios when observing lepton number violating processes. In combination with the observation of lepton flavor violating processes, we can further strengthen this argument, closing the loophole of asymmetries being stored in different lepton flavors.}

\FullConference{The European Physical Society Conference on High Energy Physics\\
		22--29 July 2015\\
		Vienna, Austria}

\begin{document}

\section{Introduction}
Different observations point us to physics beyond the Standard Model (SM). One of them is the baryon asymmetry, quantified by the measured baryon-to-photon number density ratio \cite{Planck:2015xua}
\begin{eqnarray}
\eta_{\mathrm{B}}^{\mathrm{obs}} =  (6.09 \pm  0.06) \times 10^{-10}.
\end{eqnarray}
With $CP$ violation within the SM being too small and the Higgs mass being too heavy and thus preventing a first order phase transition \cite{Gavela}, models of baryogenesis in new physics scenarios have to be evoked. 
While in the literature a variety of successful models has been proposed for leading to the observed baryon asymmetry, it is worthwhile to think about possibilities of narrowing down these options and to proceed in the question which mechanism is realised in nature. 

Many models, like for example leptogenesis, work via the principle of creating a $(B-L)$ asymmetry above the electroweak (EW) scale which is then converted via non-perturbative $(B+L)$ violating sphaleron processes above and at the EW scale to the necessary baryon asymmetry while satisfying the three Sakharov conditions \cite{Sakharov}. This principle can be also seen the other way round. Observing any $(B-L)$ violating process at low scales can lead together with $(B+L)$ violating sphaleron processes to a washout of a pre-existing baryon asymmetry. This has far-reaching consequences: Observing $\Delta L = 2$ lepton number violation (LNV) can lead to the exclusion of baryogenesis models above a certain scale irrespective of the underlying, concrete mechanism. Thus, experiments probing lepton number violation (LNV) are powerful probes in falsifying models of baryogenesis. Mainly based on our previous publications \cite{Deppisch1,Deppisch2} but also \cite{Deppisch:2014hva, Harz:2015fwa}, we will review the significant impact a possible observation of neutrinoless double beta ($0\nu \beta \beta$) decay could have as well as a signal of LNV at the LHC. We also comment on how an observation of lepton flavour violation (LFV) can strengthen our argumentation, closing the loophole of an asymmetry being stored in different flavours.

\section{Probing Lepton Number Violation with Neutrinoless Double Beta Decay}
\begin{figure}
\centering
\includegraphics[clip,width=0.24\linewidth]{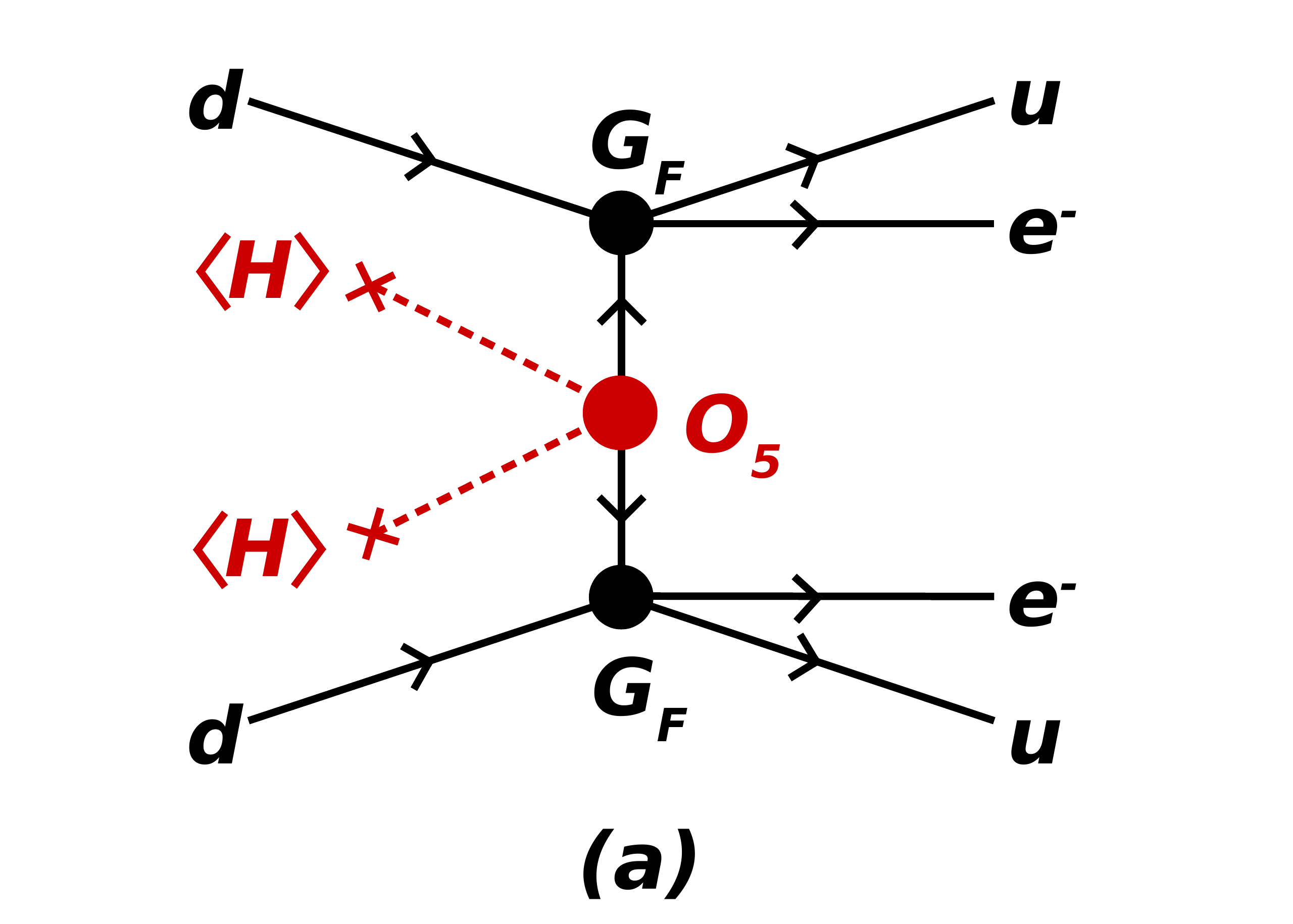}
\includegraphics[clip,width=0.24\linewidth]{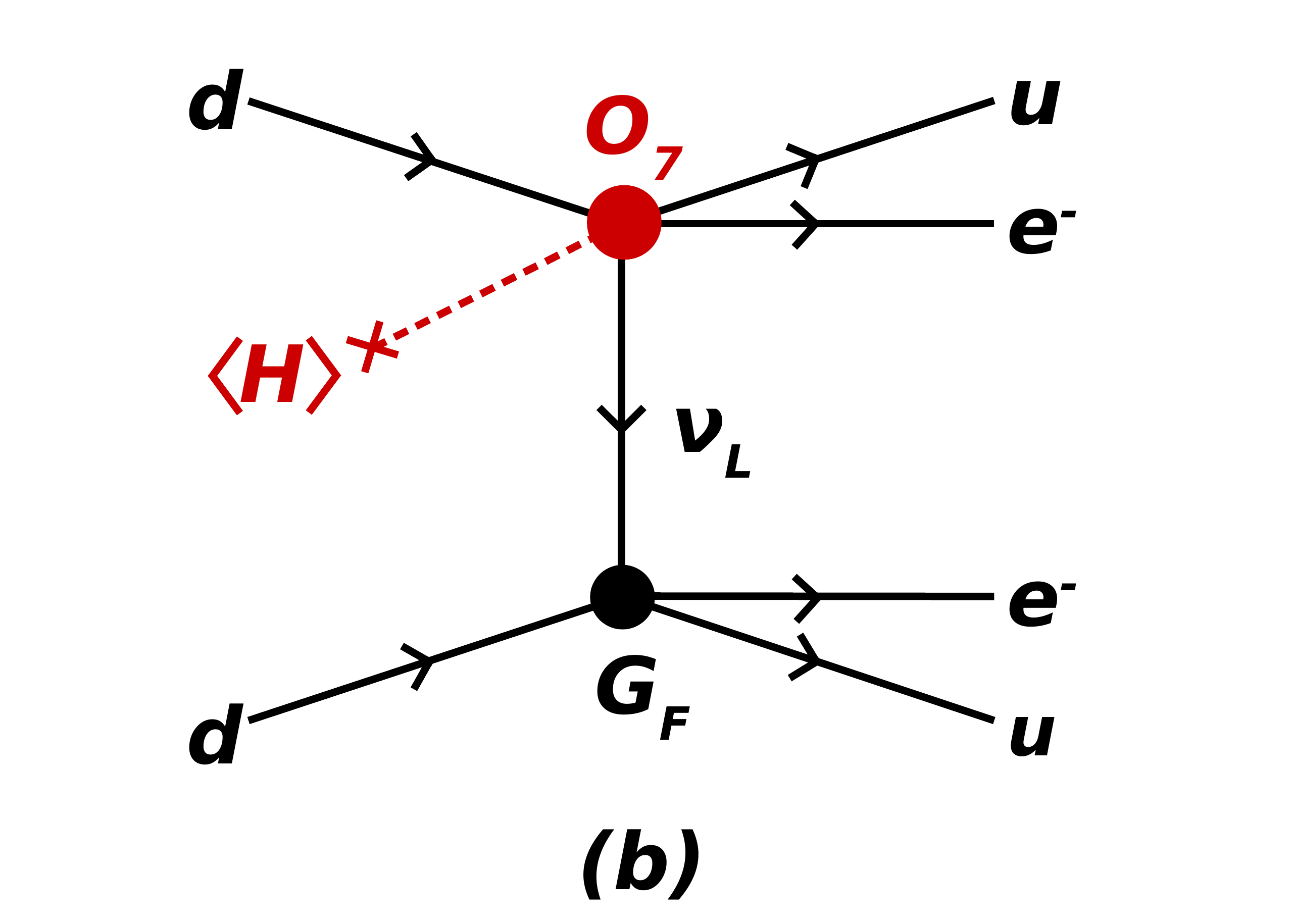}
\includegraphics[clip,width=0.24\linewidth]{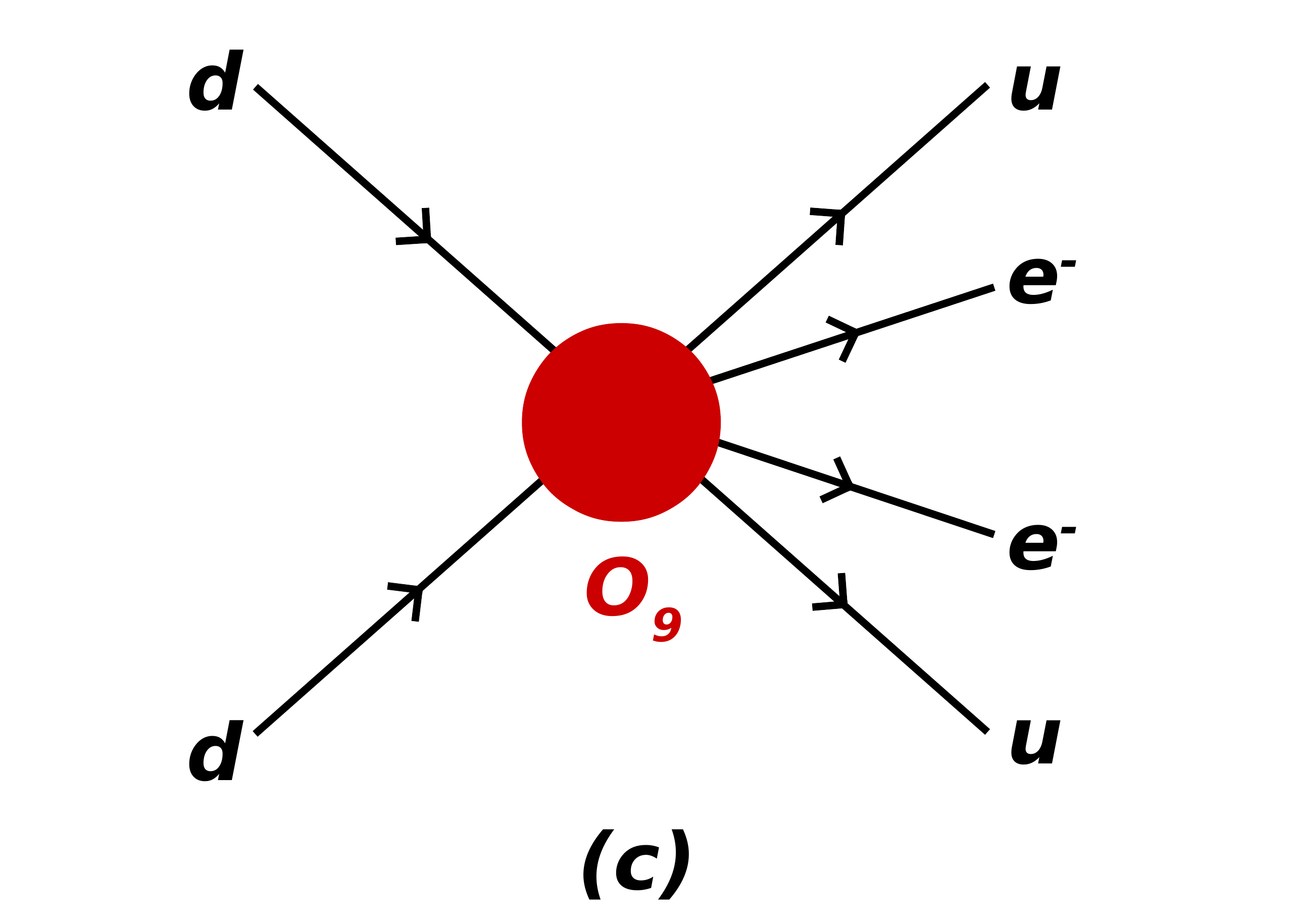}
\includegraphics[clip,width=0.24\linewidth]{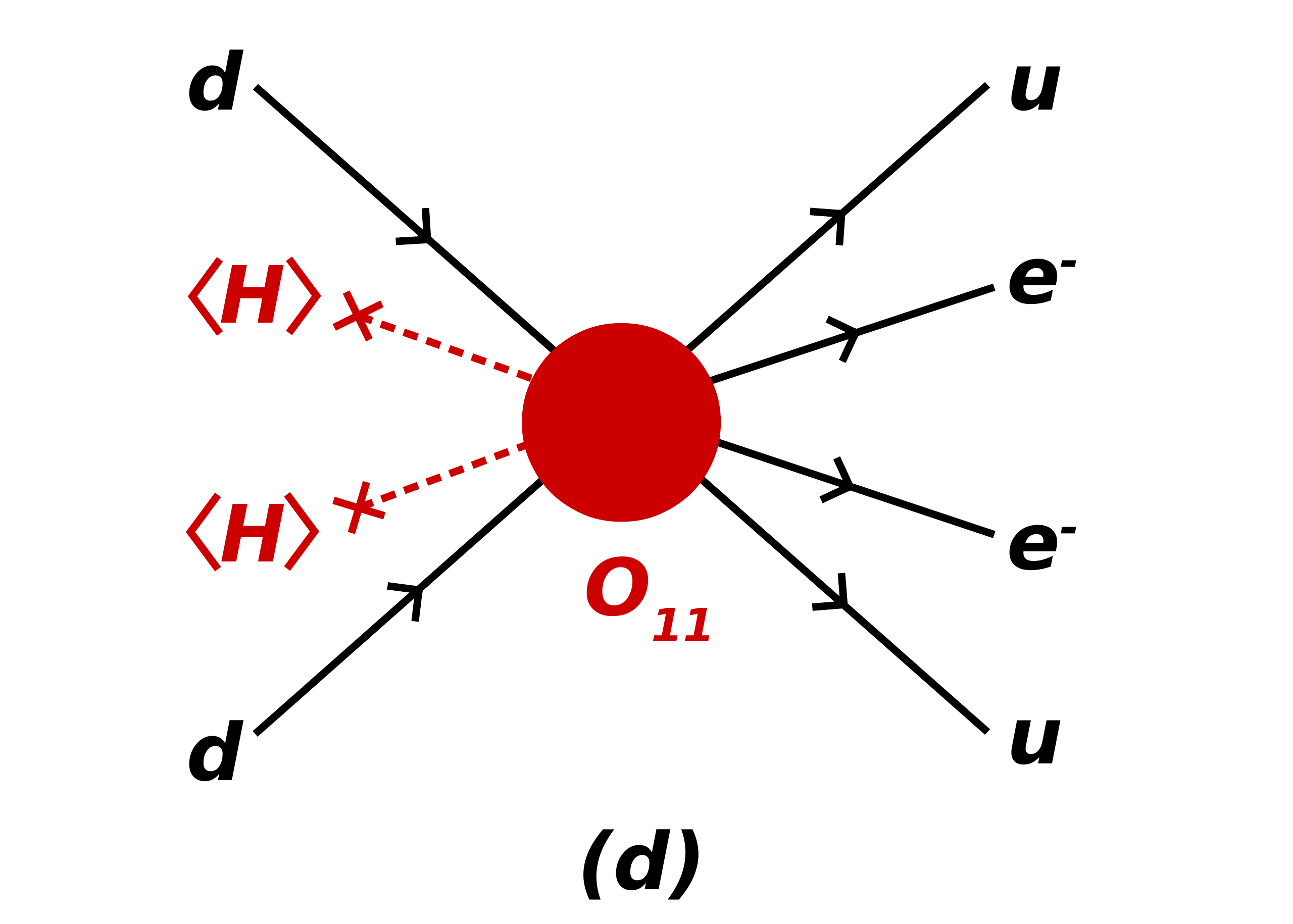}
\caption[]{Contributions to $0\nu\beta\beta$ decay generated by the operators $\mathcal{O}_5$ (a), $\mathcal{O}_7$ (b), $\mathcal{O}_9$ (c) and $\mathcal{O}_{11}$ (d), as given in Eq.~\ref{eq:operators}, in terms of effective vertices, pointlike at the nuclear Fermi momentum scale.}
\label{Fig:0vbb}  
\end{figure}

$0\nu \beta \beta$ decay, the transition of two neutrons into two protons and two electrons without any emission of anti-neutrinos is a sensitive probe of LNV. Different classes of diagrams can contribute in principle to this process as exemplified in Fig.~\ref{Fig:0vbb}: The long-range parts (a) - (b), including the standard mass mechanism (a), as well as the short-range parts (c) - (d).
With the vertices being pointlike at the Fermi scale, the long range parts can be expressed in terms of effective couplings $\epsilon_{\beta}^{\alpha}$, leading to the following general Lagrangian \cite{Deppisch:2012nb}:
\begin{equation}
\mathcal{L}=\frac{\mathrm{G_F}}{\sqrt{2}}(j^\mu_{V-A}J^\dagger_{V-A,\mu} + \sum_{\alpha \beta} \epsilon_\alpha^\beta j_\beta J^\dagger_\alpha),
\label{Eq:longrangeLagrangian}
\end{equation}
with $J^\dagger_\alpha = \bar{u} \mathcal{O}_\alpha d$ being the hadronic and $j_\beta = \bar{e} \mathcal{O}_\beta \nu$ the leptonic current and the operators $\mathcal{O}_{\alpha,\beta}$ defined as 
\begin{eqnarray}
\mathcal{O}_{V\pm A}=\gamma^\mu (1 \pm \gamma_5),\qquad
\mathcal{O}_{S\pm P}=(1 \pm \gamma_5),\qquad
\mathcal{O}_{T_{R,L}}=\frac{i}{2} [\gamma_\mu, \gamma_\nu] \gamma^\mu (1 \pm \gamma_5).
\label{Eq:longrangepart}
\end{eqnarray}
The first term in Eq.~\ref{Eq:longrangeLagrangian} leads to the standard light Majorana neutrino exchange.
Assuming one operator being dominant at a time, one can set limits on the effective couplings from the bounds on the $0\nu \beta \beta$ decay half life $T_{1/2}$ \cite{Deppisch:2012nb},
\begin{equation}
\label{eq:halflife}
 T_{1/2}^{-1} = |\epsilon_\alpha^\beta|^2 G_i |M_i|^2,
\end{equation}
where $G_i$ stands for the phase space factor and $M_i$ for the specific nuclear matrix element. Currently, no sign for this process has been found and experiments set only limits on the $0 \nu \beta \beta$ half life in different isotopes $T_{1/2}^{\mathrm{Xe}} > (1.9-1.1) \times 10^{25}$~y \cite{Albert:2014awa, Gando:2012zm}  and  $T_{1/2}^{\mathrm{Ge}} > 2.1 \times 10^{25}$~y \cite{Agostini:2013mzu}. With these current limits we arrive at the bounds on the effective couplings as given in Tab.~\ref{tab:limits}.
\begin{table}[t]
\centering
\begin{tabular}{ccccccc}
\hline
Isotope & $|\epsilon^{V+A}_{V-A}|$ &  $|\epsilon^{V+A}_{V+A}|$ & 
          $|\epsilon^{S+P}_{S-P}|$ &  $|\epsilon^{S+P}_{S+P}|$ & 
          $|\epsilon^{TR}_{TL}|$   &  $|\epsilon^{TR}_{TR}|$   \\
\hline
$^{76}$Ge & $3.3 \cdot 10^{-9}$  & $5.9 \cdot 10^{-7}$   & 
          $1.0 \cdot 10^{-8}$    & $1.0 \cdot 10^{-8}$   & 
          $6.4 \cdot 10^{-10}$   & $1.0\cdot 10^{-9}$    \\
$^{136}$Xe & $2.6 \cdot 10^{-9}$  & $5.1 \cdot 10^{-7}$   & 
          $6.2 \cdot 10^{-9}$    & $6.2 \cdot 10^{-9}$   & 
          $4.4 \cdot 10^{-10}$   & $7.4\cdot 10^{-10}$    \\
\hline
\end{tabular}
\caption[]{Limits on effective long-range $(B-L)$-violating couplings, taken from Ref.\ \cite{Deppisch:2012nb} and updated to the current limits of $T^{^{76}\mathrm{Ge}}_{1/2} > 2.1 \times 10^{25}$~y~\cite{Agostini:2013mzu} and $T^{^{136}\mathrm{Xe}}_{1/2} > 1.9 \times 10^{25}$~y~\cite{Gando:2012zm} with 90\% C.~L. limits. It is assumed that only one $\epsilon$ is different
from zero at a time.}
\label{tab:limits}
\end{table}
This implies, however, as well that as soon as $0 \nu \beta \beta$ decay is observed, the corresponding effective couplings $\epsilon_\alpha^\beta$ can be determined assuming one of them being different from zero at a time. Short range operators can be treated in a similar way. For further details we refer the reader to Ref.~\cite{Deppisch:2012nb} and references therein. 

Ref.~\cite{deGouvea:2007xp} lists all 129 possible $ \Delta L = 2$ effective operators based on the SM content and gauge structure up to dimension 11. In the following, we will discuss four of them, which can lead to the contributions shown in Fig.~\ref{Fig:0vbb}, 
\begin{eqnarray}
	\label{eq:operators}
	\mathcal{O}_5    &=& (L^i L^j) H^k H^l \epsilon_{ik} \epsilon_{jl},         \nonumber\\ 
	\mathcal{O}_7    &=& (L^i d^c) (\bar{e^c} \bar{u^c}) H^j \epsilon_{ij},     \nonumber\\ 
	\mathcal{O}_9    &=&(L^i L^j) (\bar{Q}_i \bar{u^c}) (\bar{Q}_j \bar{u^c}), \nonumber\\ 
	\mathcal{O}_{11} &=& (L^i L^j) (Q_k d^c) (Q_l d^c) H_m \bar{H_i} 
	\epsilon_{jk}\epsilon_{lm}. 
\end{eqnarray}
Hereby, $L = (\nu_L, e_L)^T$, $Q = (u_L, d_L)^T$, $H = (H^+, H^0)^T$, $e^c$, $u^c$ and $d^c$ are the SM fields. The fermions are written in terms of left-handed two-component fields. The bracketing denotes the chosen Lorentz contraction suppressing possible flavor or colour structures. The scales of the operators, as given in Eq.~\ref{eq:operators}, can be connected with the effective couplings as follows \cite{Deppisch2}
\begin{eqnarray}
\label{eq:epsilons}
	                m_e \epsilon_5          = \frac{g^2 v^2}{\Lambda_5},   \qquad
	\frac{G_F \epsilon_7}{\sqrt{2}}         = \frac{g^3 v}{2 \Lambda_7^3}, \qquad
	\frac{G_F^2 \epsilon_{\{9,11\}}}{2 m_p} = \{\frac{g^4}{\Lambda_9^5},
	                                            \frac{g^6 v^2}{\Lambda_{11}^7} \}.
\end{eqnarray}
For the Weinberg operator, the effective coupling is given by $\epsilon_5 = m_{ee}/m_e$, with $m_{ee}$ being the effective $0 \nu \beta \beta$ mass and $m_e$ the electron mass. The higher dimensional couplings are normalised with respect to the Fermi coupling $G_F$ and the proton mass $m_P$. Further, $v=174~\mathrm{GeV}$ denotes the Higgs vacuum expectation value, and $g$ a generic, average coupling constant to illustrate the scaling in an ultraviolet (UV) completed theory, set to one in the following. Given Eq.~\ref{eq:halflife} and Eq.~\ref{eq:epsilons}, a measured half life of $0 \nu \beta \beta$ decay is directly linked to the scale of the operator assuming one is dominant at a time. Table \ref{tab:LNVoverview} summarises the scales $\Lambda_D^0$ for the current sensitivity of $T_{1/2}=2.1 \times 10^{25}$~y.
\begin{table}[t!]
\begin{center}
\begin{tabular}{llll}
\hline
$\mathcal{O}_D$    & ~~$\lambda^0_D$ [GeV]  & ~~$\hat{\lambda}^0_D$ [GeV]  & ~~$\Lambda^0_D$ [GeV]\\
\hline
$\mathcal{O}_5$    & ~~$9.2 \times 10^{10}$ & ~~$1.5 \times 10^{12}$ &~~$9.1\times 10^{13}$ \\
$\mathcal{O}_7$    & ~~$1.2 \times 10^{2}$  & ~~$2.8 \times 10^{2}$ &~~$2.6\times 10^{4}$  \\
$\mathcal{O}_9$    & ~~$4.3 \times 10^{1}$  & ~~$1.7 \times 10^{2}$ &~~$2.1\times 10^{3}$  \\
$\mathcal{O}_{11}$ & ~~$7.8 \times 10^{1}$  & ~~$1.7 \times 10^{2}$ &~~$1.0\times 10^{3}$  \\
\hline
\end{tabular}
\caption[]{Operator scale $\Lambda^0_D$ and minimal washout scales $\lambda_D^0$, $\hat{\lambda}^0_D$ for the LNV operators in Eq.~\ref{eq:operators} and the current $0 \nu \beta \beta$ sensitivity $T_{1/2} = 2.1 \times 10^{25}$~y.}
\label{tab:LNVoverview}
\end{center}
\end{table}

\section{Lepton Asymmetry Washout}
In the following, we discuss the strength of the washout introduced by each single operator given in Eq.~\ref{eq:operators} on a pre-existing baryon asymmetry. For example, in the case of $\mathcal{O}_7$, 20 different permutations leading to $2 \to 3$ and $3 \to 2$ processes have to be taken into account. Thus, the Boltzmann equation for the net lepton asymmetry $\eta_L$, normalised to the photon density $n_\gamma$, can be expressed as \cite{Deppisch2}
\begin{eqnarray}
\label{eq:be}
	 n_\gamma H T \frac{d\eta_L}{d T} = c_D \frac{T^{2D-4}}{\Lambda_D^{2D-8}} \eta_L,
\end{eqnarray}
with $n_\gamma \approx 2 T^3 / \pi^2$, the Hubble parameter $H \approx 1.66 \sqrt{g_\ast} T^2 / \Lambda_{\mathrm{Pl}}$,  $g_\ast \approx 107$, and $\Lambda_{\mathrm{Pl}} = 1.2 \times 10^{19}~\mathrm{GeV}$.
The constant $c_D$ is derived from the calculation of the scattering amplitude and is given by $c_{\{5,7,9,11\}} = \{ 8/\pi^5, 27/(2\pi^7), 3.2\times 10^4/\pi^9, 3.9\times 10^5/\pi^{13}\}$. 

The $\Delta L = 2$ processes of each underlying operator $\mathcal{O}_D$ are in equilibrium and the washout of the asymmetry is effective when the washout rate is larger than the Hubble expansion \cite{Deppisch2}
\begin{eqnarray}
\label{eq:washout}
  \frac{\Gamma_W}{H} &\equiv \frac{c_D}{n_\gamma H}\frac{T^{2D-4}}{\Lambda_D^{2D-8}} = c_D' \frac{\Lambda_\mathrm{Pl}}{\Lambda_D}\left(\frac{T}{\Lambda_D}\right)^{2D-9} \gtrsim 1,
\end{eqnarray}
with $c_D' = \pi^2 c_D/(3.3 \sqrt{g_*}) \approx 0.3 c_D$. This is the case in the temperature interval \cite{Deppisch2}
\begin{eqnarray}
\label{eq:temp_limit}
  \Lambda_D \left( \frac{\Lambda_D}{c_D' \Lambda_\mathrm{Pl}} \right)^{\frac{1}{2D-9}} 
	\equiv \lambda_D \lesssim T \lesssim \Lambda_D.
\end{eqnarray}
Hereby, the lower limit denotes the scale above which the washout generated by the $\Delta L = 2$ operator is highly effective. Assuming the current limits on the $0 \nu \beta \beta$ half life, the corresponding values for $\lambda_D^0$ are given in Table~\ref{tab:LNVoverview}. The upper limit $\Lambda_D$ specifies up to which scale the effective operator approach is still valid, and from which scale on a UV complete theory has to be established. 

A more precise estimation can be done by directly solving the Boltzmann equation and determining above which scale a pre-existing lepton asymmetry of order one has been washed out to the observed value of baryon asymmetry. This leads to the more accurate scale $\hat{\lambda}_D^0$ \cite{Deppisch2},
\begin{eqnarray}
	\hat\lambda_D \approx 
	\left[(2D-9) \ln\left(\frac{10^{-2}}{\eta_B^\mathrm{obs}}\right) \lambda_D^{2D-9} + v^{2D-9}\right]^{\frac{1}{2D-9}}\!\!\!,
\end{eqnarray}
given as well in Table~\ref{tab:LNVoverview}. 

\begin{figure}[t!]
\centering
\includegraphics[clip,width=0.46\linewidth]{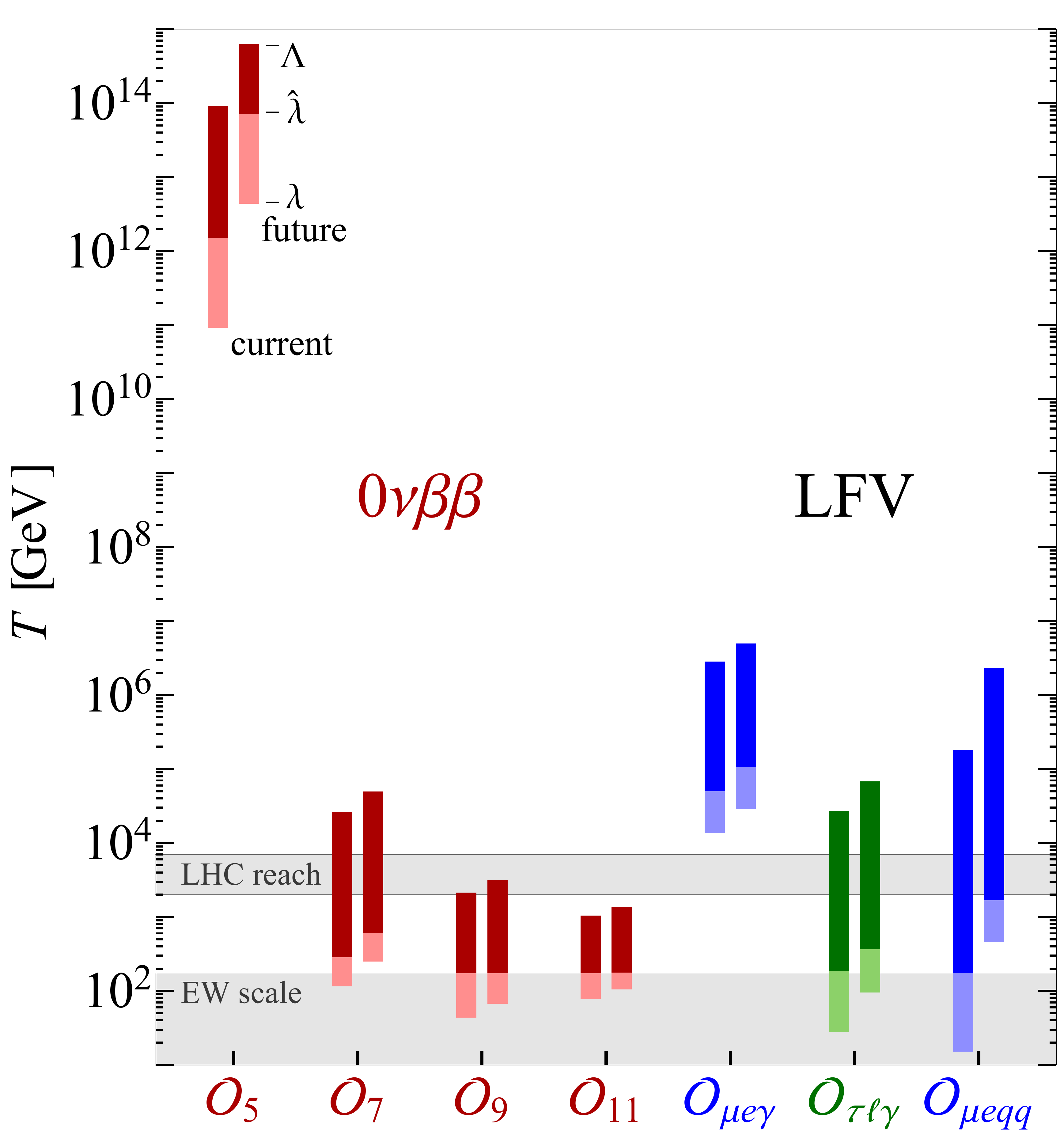}
\caption[]{Temperature intervals in which the given LNV and LFV operators are in equilibrium, defined by the operator scale $\Lambda_D$ and the minimal washout scales $\lambda_D, \hat{\lambda}_D$ as described in the text. For each operator, the left (right) bar shows the corresponding process being observed at the current (future) experimental sensitivity. For the future sensitivity we use $T_{1/2}\approx 10^{27}$~y.}
\label{Fig:ranges}  
\end{figure}

The results are summarised in Fig.~\ref{Fig:ranges}, where the current and future values for $\lambda_D$, $\hat{\lambda}_D$ and $\Lambda_D$ are depicted for each $\Delta L = 2$ operator. A striking observation is the stark difference in scales between the Weinberg operator $\mathcal{O}_5$ and the higher dimensional operators $\mathcal{O}_{7,9,11}$. As the scale $\hat{\lambda}_D$, above which a pre-existing baryon asymmetry is washed out to the observed value, lies close to the EW scale for the operators $\mathcal{O}_{7,9,11}$, observing $0 \nu \beta \beta$ via a non-standard mechanism excludes baryogenesis as a high-scale phenomenon. However, to be fair, this assumes a washout equally distributed in all flavours. As $0 \nu \beta \beta$ decay only probes the electron flavour sector, one has to ensure that all flavours are in equilibrium. 

\section{Extension to other flavours by considering Lepton Flavour Violation}
The most stringent limits on LFV are currently set on 6-dimensional $\Delta L = 0$ operators of the form $\mathcal{O}_{\ell\ell\gamma} = \mathcal{C}_{\ell\ell\gamma} \bar L_\ell \sigma^{\mu\nu} \bar{\ell^c} H F_{\mu\nu}$ and $\mathcal{O}_{\ell\ell q q} = \mathcal{C}_{\ell\ell q q} (\bar{\ell} \, \Pi_1 \ell) (\bar{q} \, \Pi_2 q)$, where $\Pi_i$ represent possible Lorentz structures ~\cite{Raidal:2008jk}, with $\ell=e,\mu,\tau$. We define the LFV operator scales $\Lambda_i$ as \cite{Deppisch2}
\begin{eqnarray}
	\mathcal{C}_{\ell\ell\gamma} = \frac{e g^3}{16\pi^2 \Lambda^2_{\ell\ell\gamma}}, \quad
	\mathcal{C}_{\ell\ell q q}    = \frac{g^2}{\Lambda^2_{\ell\ell q q}},
\end{eqnarray}
where a generic coupling $g$ as scaling of an UV complete model is introduced, set to one in our calculation. As $\mathcal{C}_{\ell\ell\gamma}$ necessarily involves an electromagnetic coupling $e$ and cannot be generated at tree level, a loop suppression factor is added as well. No correlation between the LFV and LNV operators is assumed. In a similar analysis to the above discussed LNV operators, we studied from which scale LFV is equilibrated. In Fig.~\ref{Fig:ranges}, the scale $\lambda_i^0$ is shown, which indicates above which the corresponding flavours are equilibrated and a pre-existing asymmetry is washed out. The validity of the effective operator approach is denoted by $\Lambda_i^0$. Both parameters are given for current and future sensitivities. The current values are calculated by taking $\mathrm{Br}_{\mu\to e \gamma} < 5.7\times 10^{-13}$~\cite{Adam:2013mnn}, $\mathrm{Br}_{\tau\to \ell\gamma} \lesssim 4.0\times 10^{-8}$ ($\ell=e,\mu$)~\cite{Agashe:2014kda} and the $\mu-e$ conversion rate $\mathrm{R}^\mathrm{Au}_{\mu\to e} < 7.0\times 10^{-13}$~\cite{Agashe:2014kda}, the expected sensitivities of ongoing and planned experiments are $\mathrm{Br}_{\mu\to e \gamma} \approx 6.0\times 10^{-14}$~\cite{Baldini:2013ke}, $\mathrm{Br}_{\tau\to \ell\gamma} \approx 1.0\times 10^{-9}$~\cite{Aushev:2010bq} and $\mathrm{R}^\mathrm{Al}_{\mu\to e} \approx 2.7\times 10^{-17}$~\cite{TDRComet}. 

Comparing the resulting ranges with those of the LNV operators, we encounter an overlap of $\tau \to \ell \mu$ and $\mu - e$ conversion with the higher dimensional LNV operators $\mathcal{O}_{7,9,11}$. This means when observing $0 \nu \beta \beta$ decay and LFV in $\tau \to \ell \mu$ and $\mu - e$ conversion of this scale, an asymmetry stored in a flavour other than the electron sector can be ruled out and high-scale baryogenesis models can be excluded. As visible in Fig.~\ref{Fig:ranges}, the limit on $\mu \to e \gamma$ is already too strong such that no overlap with the LNV ranges is possible. 

This demonstrates that the observation of $0 \nu \beta \beta$ decay can impose a stringent constraint on models of high-scale baryogenesis. If $0 \nu \beta \beta$ decay is observed via a non-standard mechanism, high-scale baryogenesis is in principal excluded. If LFV is observed as well, as discussed above, the possibility of an asymmetry being stored in another flavour sector can be excluded as well. Crucial at this point is the discrimination between the Weinberg operator and any non-standard mechanism leading to $0 \nu \beta \beta$ decay. Different experimental strategies exist to achieve this disentanglement. A potential discrepancy between the neutrino masses determined from cosmology and the $0 \nu \beta \beta$ half life measurement could indicate for example an underlying non-standard mechanism. Also SuperNEMO ~\cite{Pahlka:2008dw, Arnold:2010tu} is designated to discriminate the $\mathcal{O}_7$ operator from others due to its possibility to measure the energy and angular distribution of the outgoing electrons \cite{Doi:1982dn}. Further possibilities include the measurements via different isotopes \cite{Deppisch:2006hb} and the comparison between $0 \nu \beta^- \beta^-$ and $0 \nu \beta^+ \beta^+$ \cite{Hirsch:1994es}. For further details we refer the interested reader to \cite{Deppisch2} and references therein. As indicated in Fig.~\ref{Fig:ranges}, the higher dimensional operators $\mathcal{O}_9$ and $\mathcal{O}_{11}$ imply as well a potential observation of LNV at the LHC. Thus, we would like to make a short excursion on the implications of observing LNV at the LHC.

\section{Lepton Number Violation at the LHC}
\begin{figure}[t]
\centering
\begin{minipage}{0.4\textwidth}
\includegraphics[clip,width=1.0\textwidth]{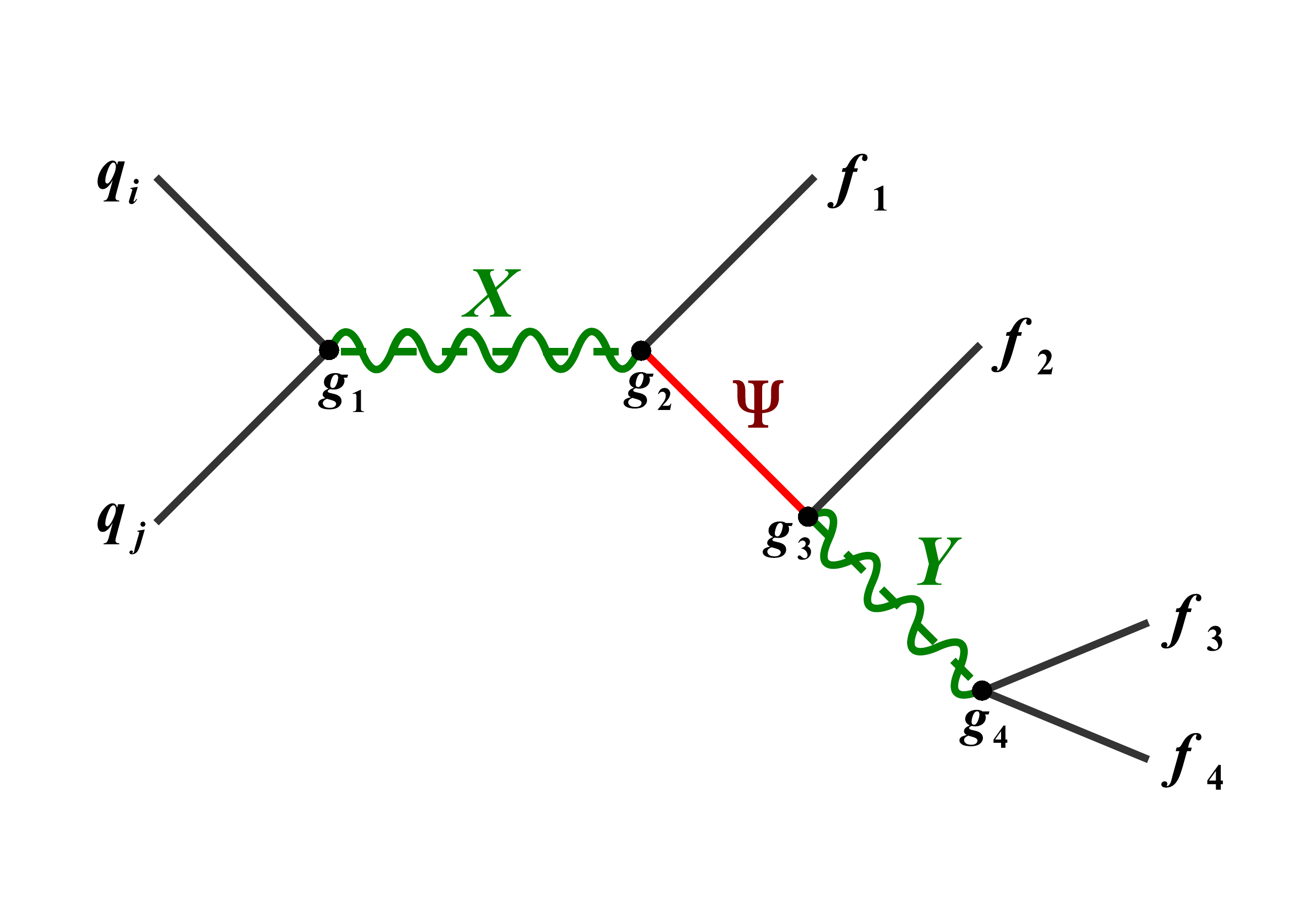}
\end{minipage}
\begin{minipage}{0.55\textwidth}
\includegraphics[clip,width=1.0\textwidth]{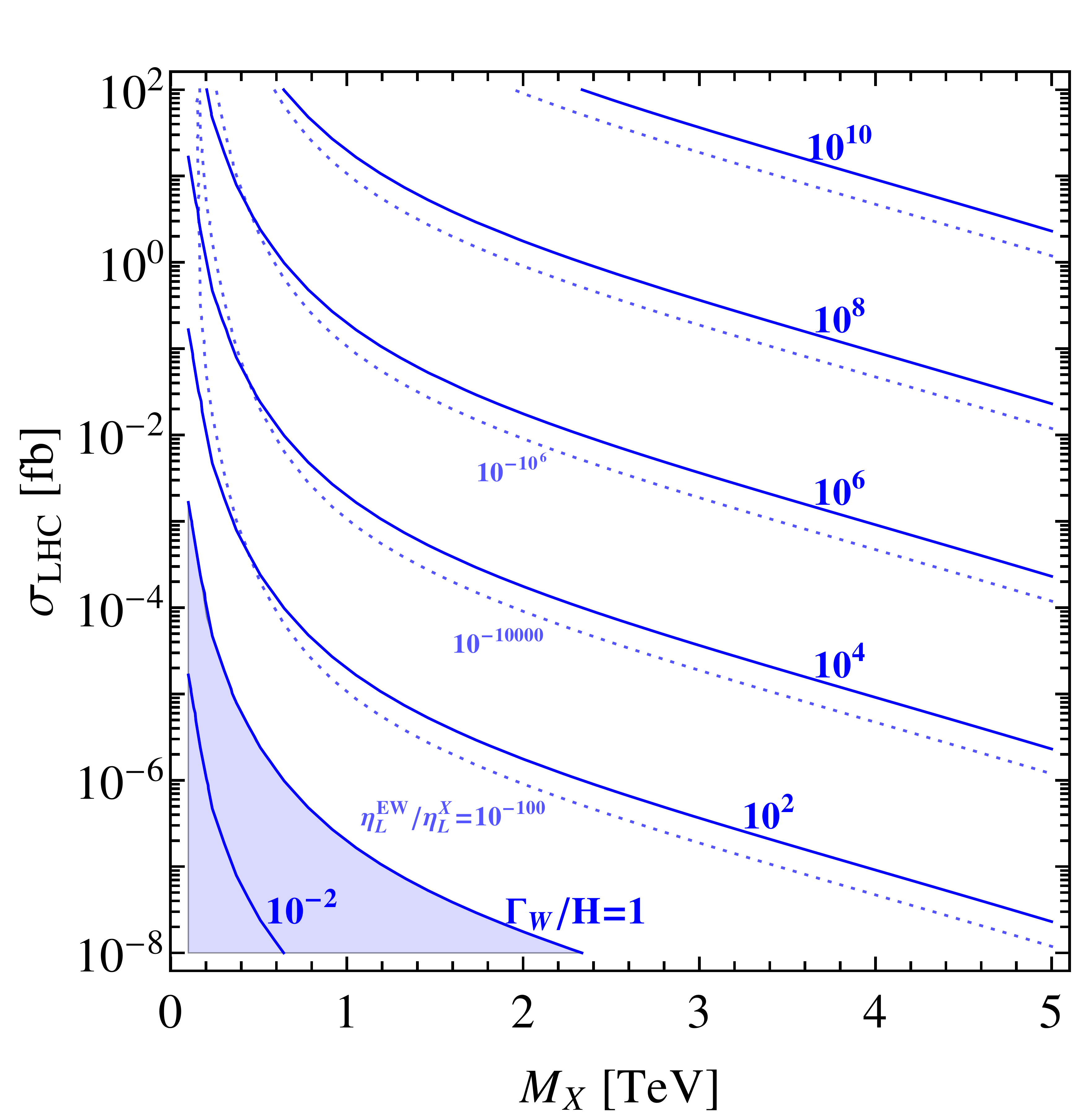}
\end{minipage}
\caption[]{Possible diagram contributing to the resonant same sign dilepton
signal $pp \to l^\pm
  l^\pm q q$ at the LHC. The particles $X$ and $Y$ indicate
different vector or scalar bosons, $\Psi$ denotes a fermion. Any two of the four fermions $f_i$ can be leptons (left). Washout rate $\Gamma_W/H$ at $T = M_X$ as a function of $M_X$
  and $\sigma_\mathrm{LHC}$ (solid blue contours). The dotted light blue lines  denote the surviving lepton asymmetry at the EW
	scale relative to its value at $M_X$, 
	$\eta_L^\mathrm{EW}/\eta_L^X$ (right).}
\label{Fig:LNVLHC}  
\end{figure}

A generic example for tree level UV completions of LNV operators has been studied in Ref.~\cite{Deppisch1}; a resonant same sign dilepton signal $pp \to l^{\pm} l^{\pm} qq$ as depicted in Fig.~\ref{Fig:LNVLHC} (left), which could be realised e.g. by a resonant $W_R$ production in L-R symmetric models. In this general approach the intermediate particles $X$ and $Y$ stand for different vector or scalar bosons and $\Psi$ denotes a fermion. Further, any two out of the four fermions $f_i$ can be leptons. Following similar arguments as above, we calculate the ratio between the washout rate and the Hubble expansion to be larger than one,
\begin{eqnarray}
\frac{\Gamma_W}{H} =  \frac{\gamma}{n_\gamma H} > 1 \qquad \mathrm{with} \qquad \gamma =  \frac{T}{32 \pi^4} \int_0^\infty d s s^{3/2} \sigma(s) K_1 (\frac{\sqrt{s}}{T}).
\end{eqnarray}
Here, $K_1$ is the 1st-order modified Bessel function of the second kind, and $\sigma(s)$ is the cross section of the $\Delta L = 2$ process. This cross section can be related to the LHC cross section $\sigma_{\mathrm{LHC}}$, factorising out a parton distribution function $f_{q_1 q_2}$, and the averaging over initial particle quantum numbers,

\begin{eqnarray}
\sigma(s) = \frac{4  \cdot 9 \cdot s_{\mathrm{LHC}}}{f_{q_1 q_2}\left( M_X / \sqrt{s_{\mathrm{LHC}}} \right)} \cdot \sigma_{\mathrm{LHC}} \cdot \delta(s - M_X^2).
\end{eqnarray}
This leads to the following expression \cite{Deppisch1},
\begin{eqnarray}
\label{eq:washout_factor_related}
  \frac{\Gamma_W}{H} &= 
  \frac{0.028}{\sqrt{g_*}}
  \frac{M_\mathrm{P}M_X^3}{T^4}
  \frac{K_1\left( M_X/T \right)}
  {f_{q_1 q_2}\left( M_X / \sqrt{s_{\mathrm{LHC}}} \right)}
  \times(s_{\mathrm{LHC}}\sigma_\mathrm{LHC}), 
\end{eqnarray}
which is independent of the branching ratios of the particle $X$ and thus valid for all coupling strengths assuming that the resonant particle $X$ and subsequent particles decay promptly at the LHC.  Evaluated at $T = M_X$ with the LHC center-of-mass energy $\sqrt{s_{\mathrm{LHC}}} = 14$~TeV, this expression only depends on the resonant mass $M_X$ and the LHC cross section $\sigma_{\mathrm{LHC}}$.

The far-reaching consequences are shown in Fig.~\ref{Fig:LNVLHC}. In blue solid lines, the washout rate is depicted. They show that a strong washout is implied when observing an LNV signal at the LHC of around $\sigma_{\mathrm{LHC}} > 10^{-2}~\mathrm{fb}$. In blue dashed lines, the strength of the washout is depicted, indicating the washout at the EW scale normalised to its value at $M_X$. In case an LNV signal at the LHC is seen, it would imply a large washout and thus render high-scale leptogenesis and baryogenesis models ineffective. Again, it should be mentioned that as a caveat it is possible that LNV is realised just in the third family only. Thus, additional measurements for $pp \to l^\pm l^\pm q q$ for either $ll = e e$, $\mu\mu$ and
$\tau\tau$, or for $e\mu$ and $e(\mu)\tau$ would testify that all lepton flavours have been equilibrated. Alternatively, the observation of LFV at low scales would allow a similar conclusion, as outlined above.

\section{Conclusions}
Searches for LNV in $0 \nu \beta \beta$ decay, at the LHC and in other contexts (e.g. in meson decays) are powerful tools to narrowing down models of baryogenesis. As demonstrated above, if $0 \nu \beta \beta$ decay was observed via a non-standard mechanism, it would point us to low-scale baryogenesis as well as a probable discovery of LNV at the LHC. If however, high-scale baryogenesis is realised in nature, no LNV is expected to be discovered at the LHC. If  $0 \nu \beta \beta$ decay was observed, its underlying mechanism is then likely to be the standard mass mechanism via the Weinberg operator and it would point us also to a high-scale origin of neutrino masses. Loop holes in this reasoning exist, like in models with hidden sectors, new symmetries or conserved charges, and we would like to refer the reader to \cite{Deppisch1,Deppisch2} for a more detailed discussion. Although those have to be considered in the specific model of baryogenesis in question, we think it is important to make theorists and experimentalists aware of the tight connection between $ 0 \nu \beta \beta$ decay and searches for LNV at the LHC with the origin of the baryon asymmetry of our Universe.

\section*{Acknowledgements}
The speaker would like to thank the organizers of EPS HEP 2015 for the opportunity to present our work and to contribute to the proceedings. Further, the authors would like to thank Martin Hirsch and Heinrich P\"as for their collaboration on the presented projects. The work of JH and WCH was supported partly by the London Centre for Terauniverse Studies, using funding from the European Research Council via the Advanced Investigator Grant 267352.


\begin{thebibliography}{99}
\bibitem{Planck:2015xua} 
  P.~A.~R.~Ade {\it et al.}  [Planck Collaboration],
  arXiv: 1502.01589 [astro-ph.CO].
  
\bibitem{Gavela}
  M.~B.~Gavela, P.~Hernandez, J.~Orloff, O.~Pene and C.~Quimbay,
  Nucl.\ Phys.\ B {\bf 430} (1994) 382.

\bibitem{Sakharov}
  A.~D.~Sakharov,
  Pisma Zh.\ Eksp.\ Teor.\ Fiz.\  {\bf 5} (1967) 32.
  
\bibitem{Deppisch1} F. F. Deppisch, J. Harz and M. Hirsch, Phys. Rev. Lett. $\mathbf{112}$ (2014), 221601.

\bibitem{Deppisch2} F. F. Deppisch, J. Harz, M. Hirsch, H. P\"{a}s and W. Huang, Phys. Rev. D$\mathbf{92}$ (2015) 036005.

\bibitem{Deppisch:2014hva}
  F.~F.~Deppisch and J.~Harz,
  Proceedings, 2nd Conference on Large Hadron Collider Physics Conference (LHCP 2014) : New York, USA, June 2-7, 2014.

\bibitem{Harz:2015fwa}
  J.~Harz, W.~C.~Huang and H.~P\"as,
  Int.\ J.\ Mod.\ Phys.\ A {\bf 30} (2015) 17,  1530045

\bibitem{Deppisch:2012nb} 
   F.~F.~Deppisch, M.~Hirsch, H.~P\"as,
   J.\ Phys.\ G {\bf 39} (2012) 124007.

\bibitem{Albert:2014awa}
  J.~B.~Albert {\it et al.}  [EXO-200 Collaboration],
  Nature {\bf 510} (2014) 229.
   
\bibitem{Gando:2012zm}
  A.~Gando {\it et al.}  [KamLAND-Zen Collaboration],
  Phys.\ Rev.\ Lett.\  {\bf 110} (2013) 6,  062502.
   
 
   
\bibitem{Agostini:2013mzu}
  M.~Agostini {\it et al.}  [GERDA Collaboration],
  Phys.\ Rev.\ Lett.\  {\bf 111} (2013) 12, 122503.


\bibitem{deGouvea:2007xp} 
  A.~de Gouvea and J.~Jenkins,
  Phys.\ Rev.\ D {\bf 77} (2008) 013008.
	
\bibitem{Raidal:2008jk}
  M.~Raidal {\it et al.},
  Eur.\ Phys.\ J.\ C {\bf 57} (2008) 13.
	
\bibitem{Adam:2013mnn}
  J.~Adam {\it et al.}  [MEG Collaboration],
  Phys.\ Rev.\ Lett.\  {\bf 110} (2013) 201801.

\bibitem{Agashe:2014kda}
  K.~A.~Olive {\it et al.}  [Particle Data Group Collaboration],
  Chin.\ Phys.\ C {\bf 38} (2014) 090001.
  
\bibitem{Baldini:2013ke}
  A.~M.~Baldini {\it et al.},
  arXiv:1301.7225 [physics.ins-det].
  
\bibitem{Aushev:2010bq}
  T.~Aushev {\it et al.},
  arXiv:1002.5012 [hep-ex].
  
 \bibitem{TDRComet}
  R. Akhmetshin {\it et al.} [COMET Collaboration],
  http://comet.kek.jp/Documents\_files/PAC-TDR-2014/PAC-Review-20141110.pdf.
\bibitem{Doi:1982dn}
  M.~Doi, T.~Kotani, H.~Nishiura and E.~Takasugi,
  Prog.\ Theor.\ Phys.\  {\bf 69} (1983) 602.
  
\bibitem{Pahlka:2008dw}
  R.~B.~Pahlka [SuperNEMO Collaboration],
  arXiv:0810.3169 [hep-ex].
 
\bibitem{Arnold:2010tu}
  R.~Arnold {\it et al.} [SuperNEMO Collaboration],
  Eur.\ Phys.\ J.\ C {\bf 70} (2010) 927.
  
\bibitem{Deppisch:2006hb}
  F.~F.~Deppisch and H.~P\"as,
  Phys.\ Rev.\ Lett.\  {\bf 98} (2007) 232501;
  V.~M.~Gehman and S.~R.~Elliott,
  J.\ Phys.\ G {\bf 34}, 667 (2007)
  [Erratum-ibid.\ G {\bf 35}, 029701 (2008)].
  
\bibitem{Hirsch:1994es}
M.~Hirsch, K.~Muto, T.~Oda and H.~Klapdor-Kleingrothaus,
\newblock Z.Phys. {\bf A347}, 151 (1994).


\end{thebibliography}
\end{document}